\documentclass[acmsmall,natbib=false]{acmart}

\AtBeginDocument{%
  }

\setcopyright{acmcopyright}

\RequirePackage[
  datamodel=acmdatamodel,
  style=acmauthoryear,backend=biber
  ]{biblatex}

\usepackage{multirow}

\begin{document}

\title{Empathetic Conversational Agents: Utilizing Neural and Physiological Signals for Enhanced Empathetic Interactions}

\author{Nastaran Saffaryazdi}
\email{zsaf419@aucklanduni.ac.nz}
\orcid{0000-0002-6082-9772}
\affiliation{%
  \institution{Empathic Computing Lab, Auckland Bioengineering Institute, University of Auckland}
  \city{Auckland}
  \country{New Zealand}
  }
\author{Tamil Selvan Gunasekaran}
\email{tg469@aucklanduni.ac.nz}
\affiliation{%
  \institution{Empathic Computing Lab, Auckland Bioengineering Institute, University of Auckland}
  \city{Auckland}
  \country{New Zealand}
}

\author{Kate Loveys}
\email{k.loveys@auckland.ac.nz }
\affiliation{%
  \institution{Department of Psychological Medicine, Faculty of Health Psychology, University of Auckland}
  \city{Adelaide}
  \country{New Zealand}
}

\author{Elizabeth Broadbent}
\email{e.broadbent@auckland.ac.nz}
\orcid{}
\affiliation{%
  \institution{Department of Psychological Medicine, Faculty of Health Psychology, University of Auckland}
  \city{Auckland}
  \country{New Zealand}
}
\author{Mark Billinghurst}
\email{mark.billinghurst@auckland.ac.nz}
\affiliation{%
  \institution{Empathic Computing Lab, Auckland Bioengineering Institute, University of Auckland}
  \city{Auckland}
  \country{New Zealand}
}

\renewcommand{\shortauthors}{Saffaryazdi et al.}

\begin{abstract}
 Conversational agents (CAs) are revolutionizing human-computer interaction by evolving from text-based chatbots to empathetic digital humans (DHs) capable of rich emotional expressions. This paper explores the integration of neural and physiological signals into the perception module of CAs to enhance empathetic interactions. By leveraging these cues, the study aims to detect emotions in real-time and generate empathetic responses and expressions. We conducted a user study where participants engaged in conversations with a DH about emotional topics. The DH responded and displayed expressions by mirroring detected emotions in real-time using neural and physiological cues. The results indicate that participants experienced stronger emotions and greater engagement during interactions with the Empathetic DH, demonstrating the effectiveness of incorporating neural and physiological signals for real-time emotion recognition. However, several challenges were identified, including recognition accuracy, emotional transition speeds, individual personality effects, and limitations in voice tone modulation. Addressing these challenges is crucial for further refining Empathetic DHs and fostering meaningful connections between humans and artificial entities. Overall, this research advances human-agent interaction and highlights the potential of real-time neural and physiological emotion recognition in creating empathetic DHs.

\end{abstract}



\received{24 December 2024}

\maketitle

\section{Introduction}
Conversational agents (CAs) are drastically altering the ways we engage with technology and manage our everyday lives in our rapidly digitizing world.  A conversational agent in human-machine interaction refers to a desktop agent, virtual character, or robot that is equipped with the capability to engage in human-like conversations with users. For CAs, it is crucial that the machine can give appropriate responses at the content level (relevant and grammatical) as well as the emotion level (consistent emotional expression) \cite{sun2018emotional}. CAs have shown promising results in various domains, such as intelligent tutoring systems, healthcare, customer services, and virtual therapy \cite{d2007toward, yalccin2018modeling, adikari2022empathic, ring2016affectively}. However, the true potential of CAs lies in their ability to go beyond mere functionality and connect with users on an emotional level. Research suggests that CAs equipped with empathetic communication skills, emotional understanding, and adaptive behavior significantly enhance their effectiveness, intelligence, and reliability \cite{lucas2014s, gratch2007creating}. This is especially important as CAs begin to move from text-based chatbots to human-like virtual characters capable of rich facial expressions and voice feedback. 

Empathy, the ability to understand and share another person's feelings, is a crucial aspect of social relationships \cite{preston2002empathy}. It involves cognitive empathy (understanding others' perspectives), affective empathy (sharing emotions), and somatic empathy (physiological response to others' emotions) \cite{preston2002empathy}. In the context of CAs, creating an empathetic interaction involves perceiving and understanding the user's emotional state (cognitive empathy) and delivering appropriate emotional responses (affective empathy) \cite{allouch2021conversational, spitale2020towards}.

Emotion recognition techniques are vital for CAs to achieve empathetic interactions, and become empathetic digital humans (DHs). This includes recognizing emotions from various inputs, such as speech, audio tone, facial expressions, and body gestures \cite{yalccin2020empathy, nandwani2021review}. Acoustic signals, for instance, are informative for speech emotion recognition (SER), where algorithms label voice inputs with specific emotional categories \cite{zhou2018inferring, hu2022acoustically}. Facial expression recognition is another approach, allowing CAs to adapt verbal and non-verbal behavior based on user emotions \cite{khosla2015service, catania2023conversational}.

Recently, neural and physiological signals have demonstrated reliability in emotion recognition \cite{blaiech2013emotion, shu2018review, ahmed2023systematic}. However, existing research has yet to exploit these cues in the perception module of Empathetic DHs. In this paper, we propose to use neural and physiological cues in the perception module of CAs to achieve more accurate empathy responses. These modalities can be especially valuable when behavioral cues, like facial expressions or speech, are difficult to measure, like in noisy and dark environments.

The primary focus of this paper is to investigate the effectiveness of integrating neural and physiological signals into the perception module of a conversational agent (CA). Additionally, it evaluates the efficiency of physiological-based multimodal emotion recognition in enhancing the empathy of virtual conversational agents. We describe the CA's components and propose the design for the perception module to foster empathy between the CA and users. A user study is conducted to evaluate the proposed CA, and the results are presented and discussed, along with potential avenues for future research. 

The main contributions  of this paper are as follows:
\begin{itemize}
    \item Using neural and physiological signals in the perception module of a conversational agent to make an empathetic interaction.
    \item Creating a publicly available dataset of multimodal data, including facial video, recorded audio, and neural and physiological signals in interaction with empathetic and neutral conversational agents.
    \item Identifying the challenges and limitations of real-time interaction with conversational agents that use neural and physiological signals. 
\end{itemize}
\section{Related Work}

\subsection{Advancements in Affective Computing and Emotion Recognition}
Due to their accessibility and ease of use, current methods of emotion recognition predominantly employ behavioral signals such as facial expressions \cite{li2018deep, samadiani2019review, sun2020multimodal}, text and audio analysis \cite{sailunaz2018emotion}. These modalities are commonly used in many human-computer interaction (HCI) applications \cite{samadiani2019review}. Some applications, like gaming \cite{alzoubi2023detecting}, self-driving cars \cite{siam2023automatic}, online embedded systems \cite{jeong2018driver}, and social robots and conversational agents \cite{ruiz2018deep, luo2022critical, antony2020emotion}, use real-time facial expression recognition and speech analysis to understand the user's state. Although these approaches display potential in identifying basic emotions, they encounter obstacles such as cultural disparities, diversity in age and gender, and the susceptibility of behavioral modalities to manipulation by the user, thereby allowing for easy deception.\cite{hossain2019observers}. 

Researchers have increasingly incorporated physiological measures like Electroencephalography (EEG) \cite{alarcao2017emotions}, Galvanic Skin Response (GSR), heart rate variability (HRV) \cite{udovivcic2017wearable}, blood volume pressure (BVP), body temperature (BT), respiration rate (RR), etc \cite{chunawale2020human} to address these limitations for emotion recognition. Many researchers have shown that although EEG and physiological signals are weak, they can be used to reliably recognize underlying emotions  \cite{blaiech2013emotion, shu2018review}. Combining these various modalities can further improve the accuracy of emotion recognition \cite{zhu2020valence, huang2019combining}. Despite advancements in neural and physiological emotion recognition, these modalities have rarely been used in real-time emotion recognition. Their potential and limitations need to be explored more deeply for real-life scenarios such as interaction with conversational agents.  

\subsection{Empathetic Conversational Agents}
The evolution of conversational agents towards achieving empathetic interactions has seen significant advancements in the past few years. Empathetic conversational agents or affective conversational agents are designed to understand and respond to users in a way that reflects an understanding of the user's emotions, providing a more human-like interaction experience \cite{smith2011interaction, yang2019understanding}. Samrose et al. \cite{samrose2020mitigating} explored the potential of empathetic conversational agents in mitigating boredom, highlighting the importance of emotional intelligence in conversational agents to adjust user mood and enhance task performance. Sakhrani et al. \cite{sakhrani2021coral} have focused on creating empathetic agents for mental health support, emphasizing the need for agents that can conduct empathetic and free-flowing conversations to assist individuals who may be hesitant to seek help from human therapists.

Most conversational agents rely only on text emotion recognition, which is a widely studied area \cite{provoost2017embodied, luo2022critical, antony2020emotion}. For example, Zaranis et al. \cite{zaranis2021empbot} developed EmpBot, an empathetic chatbot that utilizes sentiment understanding and empathy forcing to enhance the empathetic responses of conversational agents. Ayshabi and Idicula \cite{ayshabi2021multi} introduced a model that employs a multi-resolution mechanism with multiple decoders to capture the nuances of user emotions and generate empathetic responses. This approach aims to improve the perception and expression of emotional states in dialogue systems. However, facial expressions and speech tone have more information about the user's emotion \cite{luo2022critical}. For example, Hussain et al. \cite{hussain2012towards} presented a novel architecture for affective interaction with conversational agents. They fused audio, video, and text to detect affective states. Takana et al. \cite{tanaka2017embodied} introduced a social skill training agent for people with Autism that uses audio-visual features for detecting emotional states. Similarly, Aneja et al. \cite{aneja2021understanding} introduced a conversational agent that can recognize human behavior during open-ended conversations and automatically align its responses to the conversational and expressive style of the other party by using multimodal inputs, including facial expressions and speech. Tavabi et al. \cite{tavabi2019multimodal} developed a multimodal deep neural network that integrates emotional tone in language and facial expressions to recognize opportunities for empathetic responses in human-agent interactions. This study highlights the potential of incorporating visual and auditory cues to enhance the emotional intelligence of Empathetic CAs.

Despite these advancements, a significant gap remains in integrating the neural and physiological data spectrum for emotion recognition. While multimodal approaches have begun to incorporate visual and auditory cues, the potential of neural and physiological signals (e.g., EEG, HRV, and EDA) in enhancing the empathetic capabilities of conversational agents remains largely untapped. These data sources could offer deeper insights into the user's emotional state and reveal genuine emotions reliably \cite{hossain2019observers, shu2018review}. This enhanced understanding would empower CAs to generate more nuanced and authentically empathetic responses. However, conversational agents rarely use neural and physiological modalities to recognize emotion in real-time. One exception was the work of Val-Calvo et al. \cite{val2020affective}, who used an affective robot for storytelling and created a realistic paradigm for evoking emotions. They recorded EEG, BVP, GSR, and video data and estimated emotional state in real-time subject-dependently. However, the robot did not use the recognition result to adapt its behavior. Similarly, Ghandeharioun et al. \cite{ghandeharioun2019emma} introduced EMMA, which is an emotion-aware chatbot that detects emotions in real-time based on the user's smartphone location and provides wellness suggestions for mental health according to detected emotions using location data. However, they did not consider any other source of data, especially physiological cues. Equipping conversational agents with physiological emotion recognition is crucial, especially in scenarios where recording audio and video is challenging due to environmental factors like noise, poor camera angles, or low light conditions. Moreover, audio and video alone often fail to fully capture emotional states. For example, when supporting older adults, individuals with autism, or those who face difficulties expressing emotions verbally or behaviorally, leveraging neural and physiological cues can provide a more accurate and effective understanding of their emotional needs.

\section{Methodology}
\subsection{Conversational agent design}
\begin{figure}[h!]
  \centering
  \includegraphics[scale=0.2]{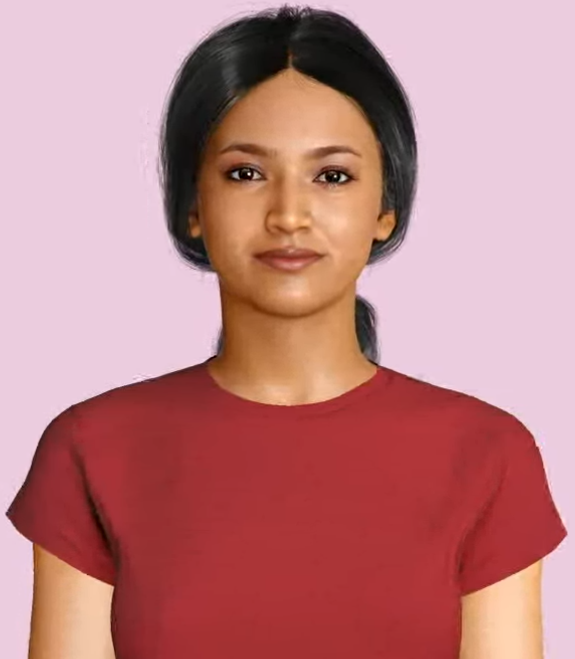}
  \caption{The Soul Machine's DH that we used in this study.}
  \Description{The upper body of a digital human which has a realistic human-like face}
  \label{fig:avatar}
\end{figure}

We used a digital human (DH) developed by Soul Machines Ltd \footnote{https://www.soulmachines.com/} (Figure \ref{fig:avatar}) to create an empathetic conversational agent. The DH was a young adult female. We disabled all of the Soul Machine features of speech and facial expression recognition so that we could implement our own. We created two different designs of DH in this research. One is a neutral DH with a simple perceptual module. It did not have an emotion recognition module and responds with a neutral pre-programmed answer and takes appropriate action based on the user's input and dialog flow. The other is an empathetic DH whose perceptual module is based on neural and physiological signals. It recognizes emotions in real-time and creates the appropriate expression by mirroring detected emotion. It chooses the appropriate response from a dataset of responses that reflect and appreciate the user's detected emotions. We used a female DH whose appearance and voice were identical in both DHs.

\subsection{Acquiring human emotional responses}
\begin{figure}[h!]
  \centering
  \includegraphics[scale=0.25]{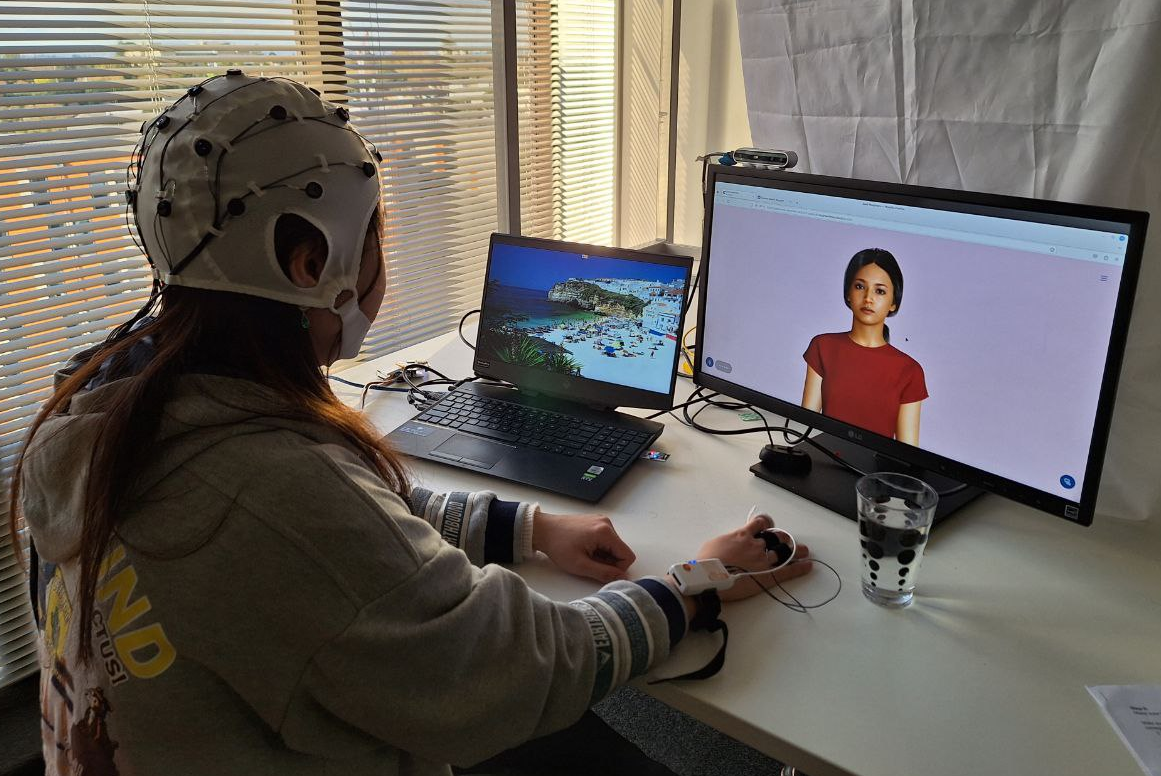}
  \caption{The experimental setup and the devices used for data collection.}
  \Description{A person sat on a chair an wore an OpenBCI EEG cap and a shimmer sensor, looking at a monitor which display a digital human. Also, a laptop displaying an image is on the table beside the monitor.}
  \label{fig:experiment_setup}
\end{figure}

In this study, we employed the Octopus-Sensing software suite \cite{saffaryazdi2022octopus} to gather data from multiple sensors and perform real-time monitoring simultaneously across all of them. The data acquisition included facial video, audio, Electroencephalography (EEG), electrodermal activity (EDA), and photoplethysmography (PPG). The EEG was captured by the OpenBCI EEG cap \footnote{https://openbci.com}, while the EDA and PPG signals were captured using the Shimmer3 sensor \footnote{https://shimmersensing.com}. Participants began by providing informed consent and receiving instructions. They were then seated in front of a monitor and laptop and fitted with an OpenBCI cap and Shimmer sensor in their non-dominant hand. Participants were alone during the data collection period in a quiet, controlled laboratory setting. The experimental setup and the devices used for data collection are shown in Figure \ref{fig:experiment_setup}. The acquired EEG, PPG, and EDA signals were primarily used for real-time emotion recognition. Although audio and video were not used in this paper, they were collected for potential future research.

\subsection{Architecture} \label{section:architecture}
\begin{figure}[h!]
  \centering
  \includegraphics[scale=0.4]{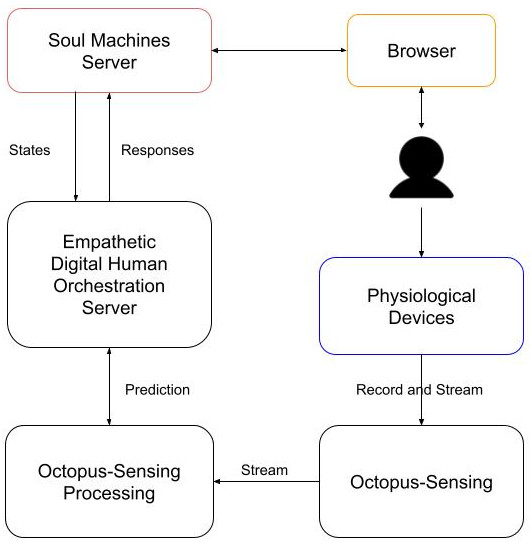}
  \caption{The overall architecture of the Embodied Conversational Agent (ECA).}
  \Description{A diagram of various components in this research and their connections}
  \label{fig:architecture}
\end{figure}

We developed an orchestration server, which acts as the Soul Machine orchestration layer, to perform several functions within the empathetic DH. The server's main tasks included understanding the user's intent, selecting the most suitable pre-written response from a database of responses, provided in the supplementary materials, and generating appropriate facial expressions. It controls the dialog flow and creates responses based on the detected emotions through a real-time emotion recognition system. The orchestration server is also responsible for regulating the expressions of the digital human based on the user's real-time emotional state. Figure \ref{fig:architecture} shows the overall architecture of the empathetic DH that works based on physiological responses.

\raggedbottom

We customized the appearance of the DH and disabled all emotion recognition features available through the Soul Machines web browser. Instead, we employed our own multimodal emotion recognition system and integrated it with the DH using an orchestration server. In the empathetic condition, the real-time emotion recognition system detected emotions every 5 seconds using a 20-second sliding window with overlap. The DH updated its facial expressions every 5 seconds based on these detections. Additionally, during interactions, the DH responded according to the most frequently detected emotion during the user's speech.

In this study, we used the dataset from prior research \cite{saffaryazdi2024exploring} to train emotion recognition models. This dataset includes EEG, PPG, and EDA data collected from 16 participants during dyadic conversations conducted in both remote and face-to-face settings, with each condition consisting of 8 trials. Due to technical issues, particularly with EEG signal recording in this dataset, some data was missing. Consequently, we had 240 trials to train the EDA and PPG models and 204 trials to train the EEG model. Since this datsets labels are based on dimensional arousal-valence model and earlier research explored that dimensional emotion models are more sufficient, reliable, and more accurate for self-assessments in explaining emotions \cite{Eerola2011comparison, lichtenstein2008comparing}, and dataset labels are based on arousal-valence model, we used a 2D-dimensional emotion model, and the results of predictions are based on arousal and valence. We mapped 5 levels of arousal and valence levels to low(-1) and high(1) values based on \ref{eq:binary_arousal}. 

We followed the same methodology as described in \cite{saffaryazdi2024exploring} to preprocess data, extract features, and train emotion recognition models using each signal. This approach has proven effective in recognizing emotions during conversations. We extracted features listed in Table \ref{tab:signal_statistical_features} and used the Random Forest Classifier (RFC), previously employed in emotion classification research \cite{menezes2017towards, chaparro2018emotion}, to train the models for real-time emotion recognition.

\begin{equation}\label{eq:binary_arousal}
Arousal = \begin{cases} 
            -1 & if ArousalRate <= 3 \\
            1 & otherwise
            \end{cases} 
\end{equation}
\begin{displaymath}
ArousalRate  = \{1, 2, 3, 4, 5 \}
\end{displaymath}

\begin{equation}\label{eq:binary_valence}
Valence = \begin{cases} 
            -1 & if ValenceRate > 3 \\
            1 & otherwise
            \end{cases}
\end{equation}
\begin{displaymath}
ValenceRate  = \{1, 2, 3, 4, 5 \}
\end{displaymath}

Eevery 5 seconds, we obtained the last 20 seconds of data from each signal, preprocessed them, extracted features from them, and fed them into the trained models to recognize emotions based on arousal and valence levels. We considered 20 seconds because the heartpy library that we used for extracting features from the PPG signal needed at least 20 seconds of data to extract HRV features. 
We used the weighted decision fusion method shown in Equation \ref{eq:fusion} to combine the recognition results from neural and physiological modalities and detect the emotional state. Equal weights were considered for detecting valence levels for all modalities' predictions. However, we considered a higher weight for EDA and PPG, which was 2 compared to EEG was 1 for arousal detection because PPG and EDA signals have been shown to have greater reliability in detecting arousal levels \cite{tarnowski2018combined, kolodziej2019electrodermal, udovivcic2017wearable}. The Equation \ref{eq:fusion_result} shows how we mapped the result of fusion to -1 and 1 values. The empathetic DH expressed emotions based on the result of the fusion method, primarily by mirroring the emotional state as listed in Table \ref{tab:reflected_expressions}.

\begin{equation}\label{eq:fusion}
  p_{o}^{x} = a \times p_{EEG}^{x} + b \times p_{GSR}^{x} + c \times p_{PPG}^{x}
\end{equation}
\begin{displaymath}
  x \in [Arousal, Valence]
\end{displaymath}
\begin{displaymath}
  a = b = c = 1    for Valence 
\end{displaymath}
\begin{displaymath}
  a = 1, b = c = 2 for Arousal
\end{displaymath}

\begin{equation}\label{eq:fusion_result}
Fusion = \begin{cases} 
            -1 & if p_{o}^{x} < 0 \\
            1 & otherwise
            \end{cases}
\end{equation}
\begin{displaymath}
x \in [Arousal, Valence]
\end{displaymath}

To map the result of predictions to the avatar's expressions, we used Table \ref{tab:reflected_expressions}.
There are not many differences between various positive expressions (with various intensity or arousal levels) in the avatar's expressions. So we considered all as happiness which means mirroring positive emotions with happiness expression. For expressions during answering, Soul machines provide happiness strong and slight, which exactly matches the arousal level. For negative emotions, including anger or fear, the soul machine expressions were quite similar, which made the mapping easier. To respond to negative emotions, the avatar used compassionate talking.

\begin{table}[]
\centering
\caption{Defining the Empathetic DH expressions according to the result of the fusion method}
\label{tab:reflected_expressions}
\begin{tabular}{|c|c|c|c|}
\hline
\multicolumn{1}{|l|}{\textbf{Arousal}} & \multicolumn{1}{l|}{\textbf{Valence}} & \multicolumn{1}{l|}{\textbf{state}} & \multicolumn{1}{l|}{\textbf{DH Expression}} \\ \hline
1                                      & 1                                     & HAHV                                & Strong Happiness                                    \\ \hline
1                                      & -1                                    & HALV                                & Anger/Scaredness                             \\ \hline
-1                                     & 1                                     & LAHV                                & Slight Happiness                                    \\ \hline
-1                                     & -1                                    & LALV                                & Sadness                                      \\ \hline
\end{tabular}
\end{table}

\section{User Study Design}
We conducted a within-group user study to investigate the empathy between a human and a DH when the DH behaves emphatically compared to when the DH behaves in a neutral way. The study procedure was approved by and performed in accordance with the guidelines and regulations of the University of Auckland Human Participants Ethics Committee on March 2, 2020 (reference number: 023799).

\subsection{Procedure}
We recruited 23 participants aged between 21 and 44 years old ($\mu=30, \sigma=6$). The participants were university students and staff, and they were offered a \$40 voucher as an incentive for participating in the study.

All participants had a set of conversations with both DHs and talked about different emotional topics in a random order after providing written informed consent. In the neutral condition, the DH asked questions, responded neutrally, and showed no expression when the participant was talking. In the empathetic condition, the DH reflected the recognized emotion in its expressions and responses. The empathetic DH expressed its feelings about the image, asked questions, and responded in an empathetic way according to the participant's response.

We followed the approach described in \cite{saffaryazdi2024exploring} to generate emotional conversations but replaced the human interviewer with the DH. We selected eight images from the IAPS image set \cite{lang2007international} as the topic of conversation, two from each of the following categories: High Arousal-High Valence (HAHV), High Arousal-Low Valence (HALV), Low Arousal-High Valence (LAHV), and Low Arousal-Low Valence (LALV). We sorted the dataset according to arousal and valence level, ignored images with high negative subjects like suicide or violence, and chose images with the highest and lowest arousal and valence levels with popular subjects to discuss. After displaying an image from the IAPS dataset, the DH started conversing with the participant about their feelings about the image and related experiences. 

\subsection{Emotion and Empathy Self-report Data}
After each conversation topic, all participants were asked to complete a self-report questionnaire which designed according to the definition of emotion and empathy and previous studies \cite{koelstra2011deap}\cite{witchel2016complex}\cite{jauniaux2020emotion}. This data was used to assess real-time emotion recognition and the level of empathy between the participants and the digital human agent. In addition, we used the self-report data to evaluate whether the interaction with the digital human could evoke emotions in the participants.

The questionnaire consisted of 8 questions, categorized into three parts. The first part (questions 1 and 2) aimed to evaluate the participant's emotional state based on the 2-dimensional emotion model. We compared their responses to these questions with the results of real-time emotion recognition for each trial to assess the quality of the emotion recognition.

The second part (questions 3 to 5) were created according to the definition of empathy and previous studies to evaluate empathy \cite{witchel2016complex}\cite{jauniaux2020emotion}. They were used to evaluate the perceived level of empathy displayed by the digital human agent towards the participants, including cognitive and affective empathy.

The third part (questions 6 to 8) focused on assessing the appropriateness and timing of the digital human's emotions and expressions. These questions were designed to gauge how well the DH's emotions and expressions aligned with the participant's emotions. Table \ref{tab:self-report-questionnaire} provides an overview of the questionnaire and its related factors.

\begin{table*}[]
\centering
\scriptsize
\caption{Self-report questionnaire after each conversation. DH: Digital Human}
\label{tab:self-report-questionnaire}
\begin{tabular}{|l|l|l|}
\hline
\textbf{Category}                                                               & \textbf{Question}                                              & \textbf{Variable}    \\ \hline
\multirow{2}{*}{Emotional state}                                                & 1- How negative or positive was the emotion that you felt?     & Valence              \\ \cline{2-3} 
                                                                                & 2- What was your arousal level: Calm to Excited?               & Arousal              \\ \hline
\multirow{3}{*}{\begin{tabular}[c]{@{}l@{}}Human-Agent\\ Empathy\end{tabular}}  & 3- I felt that DH was empathetic toward me.                  & Empathic             \\ \cline{2-3} 
                                                                                & 4- DH identified what I was feeling.                         & Cognitive            \\ \cline{2-3} 
                                                                                & 5- DH reflected on my emotions.                              & Affective            \\ \hline
\multirow{3}{*}{\begin{tabular}[c]{@{}l@{}}real-time\\ Expressions\end{tabular}} & 6- The emotion that DH displayed was appropriate.            & Appropriate\_emotion \\ \cline{2-3} 
                                                                                & 7- DH displayed the best expression for the situation.       & Best\_expression     \\ \cline{2-3} 
                                                                                & 8- DH’s emotional expressions occurred at appropriate times. & Appropriate\_time    \\ \hline
\end{tabular}
\end{table*}

\subsection{Human-Agent Rapport}
To measure the participants' feelings about their interaction with the agent, we used the post-interaction questionnaire after each condition, adapted from the questionnaire used in earlier research \cite{cerekovic2014you, heerink2010assessing}. As shown in Table \ref{tab:human-agent-rapport}, This questionnaire featured statements that could be responded to on a 5-point Likert scale. This questionnaire evaluates the quality of interaction (QoI), degree of rapport (DoR), degree of liking the agent (DoL), and degree of social presence (DSP). The first three factors was derived from the perception of interaction questionnaire developed by Crekovic et al. \cite{cerekovic2014you}. They have been inspired by the study \cite{cuperman2009big}, in which the authors examine how the Big Five personality traits are expressed in mixed-sex dyadic interactions between strangers. To measure perceived interaction, they developed a "Perception of Interaction" questionnaire containing items that evaluate various aspects of participants' interaction experiences. The last factor, DSP, has been validated by Heerink et al. \cite{heerink2010assessing} with Cronbach's alpha of 0.83, indicating good reliability. Table \ref{tab:human-agent-rapport} shows the details of the human-agent rapport questionnaire.

\begin{table}[]
\caption[Human-agent Raport questions]{Human-agent Raport questionnaire and the category of each question. The categories include quality of interaction (QoI), degree of rapport (DoR), degree of liking the agent (DoL), and degree of social presence (DSP).}
\label{tab:human-agent-rapport}
\begin{tabular}{|l|c|}
\hline
\textbf{Question}                                                                                                                            & \textbf{Category} \\ \hline
\begin{tabular}[c]{@{}l@{}} The interaction with DH was smooth, \\ natural, and relaxed.                                                                                  \end{tabular}   & QoI               \\ \hline
I felt accepted and respected by DH.                                                                                                       & DoR               \\ \hline
I think DH is likable.                                                                                                                    & DoL               \\ \hline
I got along with DH pretty well.                                                                                                           & DoR               \\ \hline
I did not want to get along with DH.                                                                                                       & DoL               \\ \hline
\begin{tabular}[c]{@{}l@{}} The interaction with DH was forced, \\ awkward, and strained.                                                                                \end{tabular}  & QoI               \\ \hline
I felt uncomfortable during the interaction.                                                                                                 & QoI               \\ \hline
I felt that DH was paying attention to my mood.                                                                                            & DoR               \\ \hline
\begin{tabular}[c]{@{}l@{}}I was paying attention to the way that DH responded \\ to me and I was adapting my own behavior to it.\end{tabular} & DoR               \\ \hline
I think DH finds me likable.                                                                                                              & DoL               \\ \hline
\begin{tabular}[c]{@{}l@{}}
When interacting with DH, I felt \\ like interacting with a real person                                                                       \end{tabular}& Sp                \\ \hline
I enjoyed the interaction.                                                                                                                   & QoI               \\ \hline
I sometimes felt like DH was actually looking at me                                                                                        & Sp                \\ \hline
I would like to interact more with DH in the future.                                                                                       & DoL               \\ \hline
I can imagine DH as a living creature                                                                                                      & Sp                \\ \hline
DH often said things completely out of place.                                                                                              & QoI               \\ \hline
The interaction with DH was pleasant and interesting.                                                                                      & QoI               \\ \hline
I often realized DH is not a real living creature                                                                                          & Sp                \\ \hline
Sometimes it seemed as if DH had real feelings                                                                                             & Sp                \\ \hline
\end{tabular}
\end{table}

\section{Analysis}

\subsection{Self-report data}
Since all participants had conversations with neutral and empathetic DHs, we performed a within-group statistical analysis to see the effect of DH type on humans. Data were analyzed using Python and the ARtool package \cite{kay2016package} in R. The self-report data were based on 5-point Likert scale, so our data was analyzed using non-parametric methods. We used the aligned rank transform (ART) with repeated measures ANOVA \cite{wobbrock2011aligned} to statistically analyze non-parametric data. A set of three-way repeated measures ANOVA with ART was conducted to examine the effect of DH type (neutral or empathetic) and emotional state of conversation on empathy, emotional state, and the appropriateness of DH responses and expressions. We transferred the reported arousal and valence levels to the low(-1) and high(1) values according to Equations \ref{eq:binary_arousal} and \ref{eq:binary_valence} to reduce the complexity and increase the recognition process in real-time. Additionally, using continuous or multi-class labels can lead to data sparsity, complicating the training of robust models. Low and high labels can address this issue by consolidating data into broader categories, making the classification results more interpretable for end-users. We considered low and high for arousal and valence values as the emotional state of conversation for statistical analysis.

Similarly, to statistically analyze the effect of agent type on the human-agent report, we conducted a set of one-way repeated measures ANOVA with ART when the agent type was the condition.

\subsection{Human body responses}
To compare the effect of agent type on human body responses, we extracted some common and descriptive features from the cleaned EEG, PPG, and EDA signals as described in Table \ref{tab:signal_statistical_features}. Since these data were not normally distributed, we used a one-way repeated measures ANOVA with ART when the agent type was the condition. 

\begin{table*}[]
\scriptsize
\centering
\caption{Features extracted from the EEG, EDA, and PPG signals }
\label{tab:signal_statistical_features}
\begin{tabular}{|l|l|l|}
\hline
\textbf{Modality} & \textbf{Features}                                                                                                & \textbf{Feature Details}                                                                                                                                                                                               \\ \hline
EEG               & \begin{tabular}[c]{@{}l@{}}Power Spectral Density (PSD) \\ of EEG band powers\end{tabular}                       & \begin{tabular}[c]{@{}l@{}}psd\_theta, psd\_alpha, psd\_beta\\ psd\_gamma, psd\_delta\end{tabular}                                                                                                                     \\ \hline
PPG               & \begin{tabular}[c]{@{}l@{}}Some HRV time domain features\\ using neurokit library\end{tabular}                   & \begin{tabular}[c]{@{}l@{}}HRV\_MeanNN, HRV\_SDNN, HRV\_RMSSD,  HRV\_SDSD, HRV\_CVNN, \\ HRV\_MedianNN, HRV\_CVSD, HRV\_MadNN, HRV\_MCVNN, HRV\_IQRNN, \\ HRV\_pNN50, HRV\_pNN20, HRV\_HTI, HRV\_TINN\end{tabular} \\ \hline
EDA               & \begin{tabular}[c]{@{}l@{}}Some statistics of phasic and \\ tonic components using neurokit library\end{tabular} & \begin{tabular}[c]{@{}l@{}}mean, standard deviation, and variance of\\ phasic and tonic components and peaks\end{tabular}                                                                                              \\ \hline
\end{tabular}
\end{table*}
\section{Results}
To evaluate the agent's empathy, we assessed the three components of empathy including cognitive empathy, affective empathy and somatic empathy \cite{preston2002empathy}. First, statistically show the human-agent empathy and rapport based on self-report data (Cognitive empathy). Second, we examine the differences in human neural and physiological responses when interacting with neutral versus empathetic avatars (Somatic empathy). Finally, we assess the quality of emotion recognition and real-time expressions (Affective empathy).
\subsection{Human-Agent Empathy}
We conducted a series of one-way repeated measure ANOVA with ART to examine the empathy between humans and both DHs. The assessment of empathy was based on three reported factors: (1) the overall empathy level between the human and agent (empathy), (2) the agent's ability to identify emotions (cognitive empathy), and (3) the agent's capacity to reflect emotions (affective empathy). The results of the analysis revealed significant differences in all of these factors, overall empathy (F(1, 166) = 27, p \textless 0.001), cognitive empathy (F(1, 166) = 4.7, p \textless 0.03), and affective empathy (F(1, 166) = 5.4, p \textless 0.02) during interactions with both neutral and empathetic DHs.

\begin{figure*}[h!]
  \centering
  \includegraphics[scale=0.25]{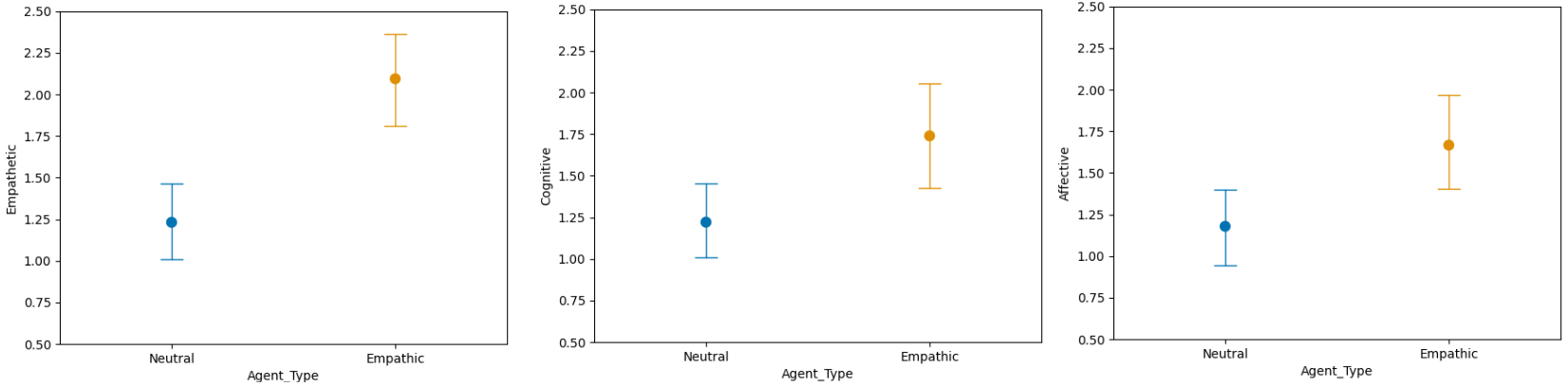}
  \caption{Effect of agent type on empathy factors including cognitive and affective empathy and overall empathy level.}
  \Description{Three point-plot that shows the effect of agent type on various empathy factors. In all point plots empathy factor was higher with the empathetic avatar}
  \label{fig:agent_empathy}
\end{figure*}

Figure \ref{fig:agent_empathy} illustrates the empathy factors based on the type of agent. It is apparent from the figure that the average scores of the empathy factors were significantly higher in conversations with the empathetic DH in comparison to the neutral DH. However, the ratings were not notably high on average. To understand the reasons behind these moderate ratings, we further evaluated the effect of participants' emotional state during the conversation on the empathy factors.

Similar to the previous section, we used a set of two-way repeated measure ANOVA with ART to examine how the emotional state affects empathy factors. The results of the ANOVA test revealed significant differences in empathy factors across various emotional arousal and valence levels (Table \ref{tab:empathy_emotion}). However, there were no significant interactions between arousal and valence levels.

\begin{figure}[h!]
  \centering
  \includegraphics[scale=0.5]{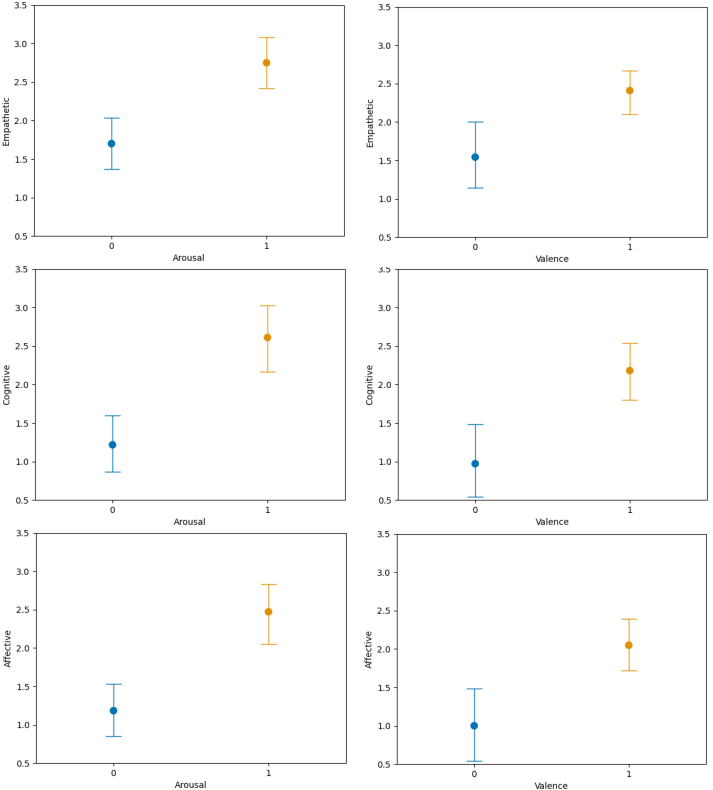}
  \caption{Effect of the emotional state of conversation topic on empathy factors.}
  \Description{Six point-plot that shows the effect of the emotional state of conversation topic on empathy factors. In all point plots, the empathy factor is higher with higher arousal and higher valence}
  \label{fig:empathy_emotion}
\end{figure}

As demonstrated in Figure \ref{fig:empathy_emotion}, the ratings for all three factors were remarkably higher during positive emotions and high arousal states. This result aligns with the result in the previous section, indicating that in high valence and arousal, because reflected emotions are more appropriate and at an appropriate time, people feel more empathy during their interaction with the DH.

\begin{table}[]
\centering
  \caption{Effect of emotional state on empathy factors in the empathetic DH.}
  \label{tab:empathy_emotion}
\begin{tabular}{|c|c|c|c|c|c|}
\hline
\textbf{Variable} & \textbf{Condition} & \textbf{F} & \textbf{Df} & \textbf{Df.res} & \textbf{p}      \\ \hline
Empathetic        & Arousal            & 17         & 1           & 84              & \textless 0.001 \\ \hline
Empathetic        & Valence            & 23         & 1           & 87              & \textless 0.001 \\ \hline
Cognitive         & Arousal            & 29         & 1           & 91              & \textless 0.001 \\ \hline
Cognitive         & Valence            & 26         & 1           & 91              & \textless 0.001 \\ \hline
Affective         & Arousal            & 29         & 1           & 90              & \textless 0.001 \\ \hline
Affective         & Valence            & 25         & 1           & 91              & \textless 0.001 \\ \hline
\end{tabular}
\end{table}

\subsection{Human-Agent Rapport}
\begin{figure}[h!]
  \centering
  \includegraphics[scale=0.3]{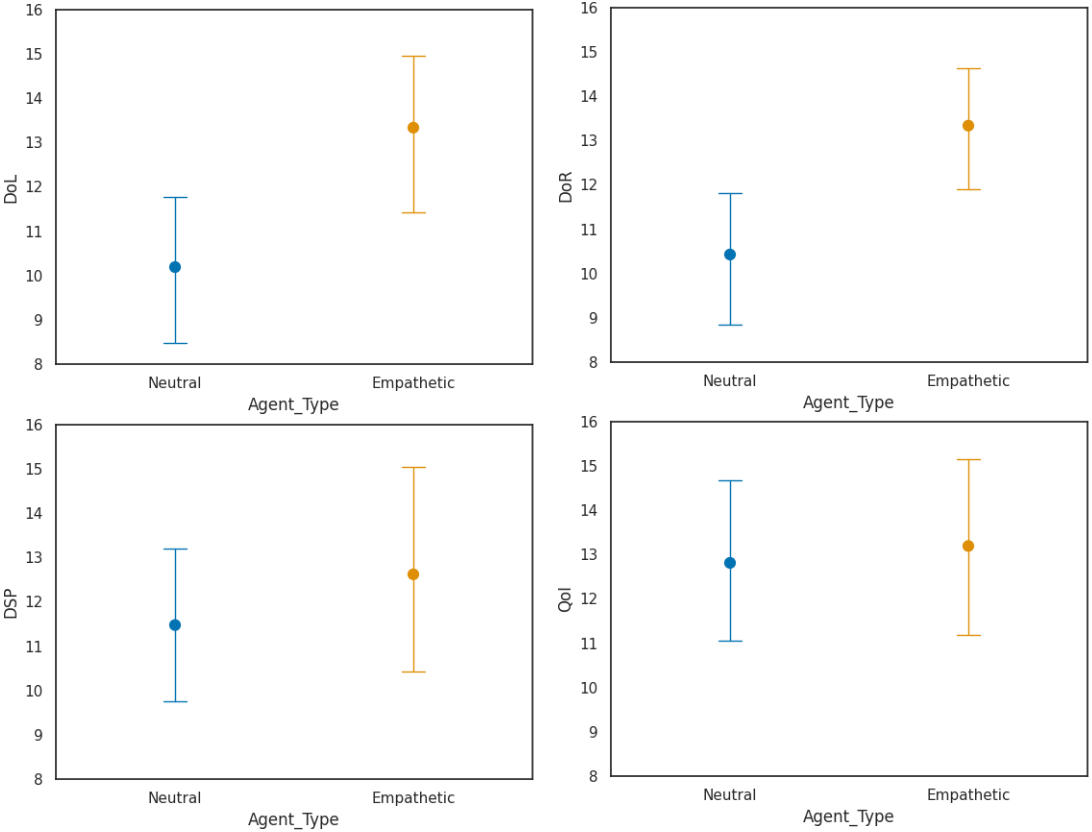}
  \caption[Point-plots of human-agent raport factors.]{Point-plots of human-agent raport factors. DoL: Degree of Liking, DOR: Degree of Rapport, DSP: Degree of Social Presence, QoI: Quality of Interaction.}
  \Description{4 point plots. x dimension is neutral and empathetic, y dimension is the rapport factor. In All Rapport is higher in interaction with empathetic avatar}
  \label{fig:post_study}
\end{figure}

A series of one-way repeated measures ANOVA was conducted to examine the effect of agent type on human-agent rapport factors. Results showed that the type of agent led to statistically significant differences in the Degree of Rapport (DoR) (F(1, 42) = 8.38, p = 0.006) and DoL (F(1, 42) = 6.64, p \textless 0.01). The other two factors, including Degree of Social Presence (DSP) and Quality of Interaction (QoI), were not significantly different. However, point plots in Figure \ref{fig:post_study} show a higher rating in the interaction with the empathetic agents. 

We also measured the percentage of participants who reported a higher level of DoR, DoL, DSP, and QoI in interaction with the empathetic agent compared to the neutral one. For all factors, a higher portion of participants reported higher levels of DoL, DoL, DSP, and QoI in interaction with the empathetic agent, which was 82\%, 86\%, 59\%, and 63\%, respectively. The higher percentage of participants who reported higher levels of DoR, DoL, DSP, and QoI in interaction with the empathetic agent suggests that the empathetic agent was able to create a more positive and engaging interaction experience for participants. This is likely due to the fact that the empathetic agent was able to understand better and respond to the emotional needs of participants.

\subsection{Physiological Responses}
\subsubsection{EEG signals}
The ANOVA test results indicated that the agent type did not significantly impact EEG features. One possible reason for this lack of significance could be the presence of considerable noise in the EEG signals caused by talking during the conversation. The high levels of noise might have made it challenging for the statistical model to identify any discernible patterns in the data. Furthermore, during a conversation, multiple activities such as decision-making, language processing, motor cortex activity, emotion processing, and memory and information processing occur. These activities are likely to be common during interactions with both types of virtual agents. As a result, any subtle differences in brain activity related to the agent type may have been overshadowed by the shared neural processes.

Future research could focus on developing more robust noise reduction methods for EEG signals to better explore the effect of agent type on brain activities. In addition, using more detailed features in the analysis could help uncover subtle variations in brain activity that might be associated with different types of virtual agents. By addressing these challenges, we can gain deeper insights into the neural mechanisms underlying human-agent interactions and their potential implications for improving virtual agent design and usability. 

\subsubsection{PPG signals}
\begin{figure}[h!]
  \centering
  \includegraphics[scale=0.3]{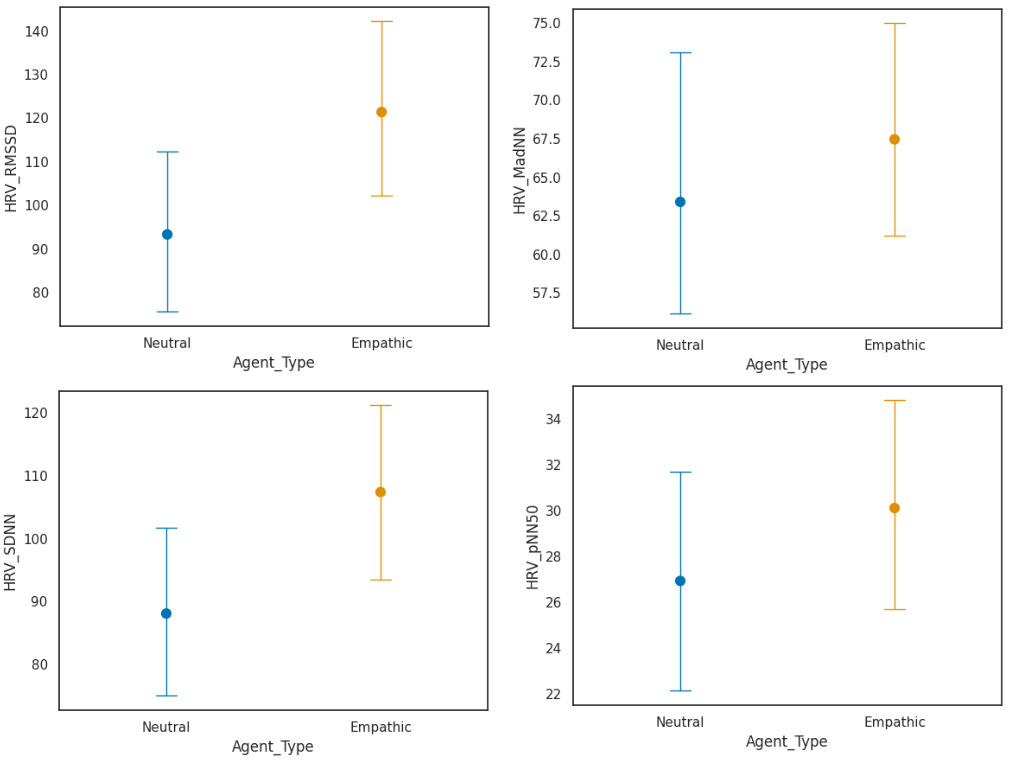}
  \caption{Point-plots of HRV features extracted from PPG signals}
  \Description{includes point plots of 4 features of HRV which is higher in interaction with empathetic agent}
  \label{fig:ppg}
\end{figure}

The ANOVA test results revealed significant differences in almost all heart rate variability (HRV) features during interactions with neutral and empathetic DH (Table \ref{tab:EDAPPG}). The point plots in Figure \ref{fig:ppg} show participants exhibited higher levels of each HRV feature in interactions with the empathetic DH, indicating greater HRV. Figure \ref{fig:ppg} shows the point plot of the most important HRV features, and the trend of others is the same.

HRV is linked to enhanced emotional well-being \cite{mather2018heart}. It is also connected to reduced levels of stress, lower anxiety \cite{chalmers2014anxiety}, and improved regulation of emotional responses \cite{appelhans2006heart} and empathy \cite{jauniaux2020emotion}. Therefore, the interactions with the empathetic agent appear to have helped individuals regulate their emotions more effectively than with the neutral DH. In addition, higher HRV is linked to increased attention \cite{appelhans2006heart}, suggesting that the empathetic agent was more engaging and effectively captured participants' attention.

\subsubsection{EDA signals}
\begin{figure*}[h!]
  \centering
  \includegraphics[scale=0.27]{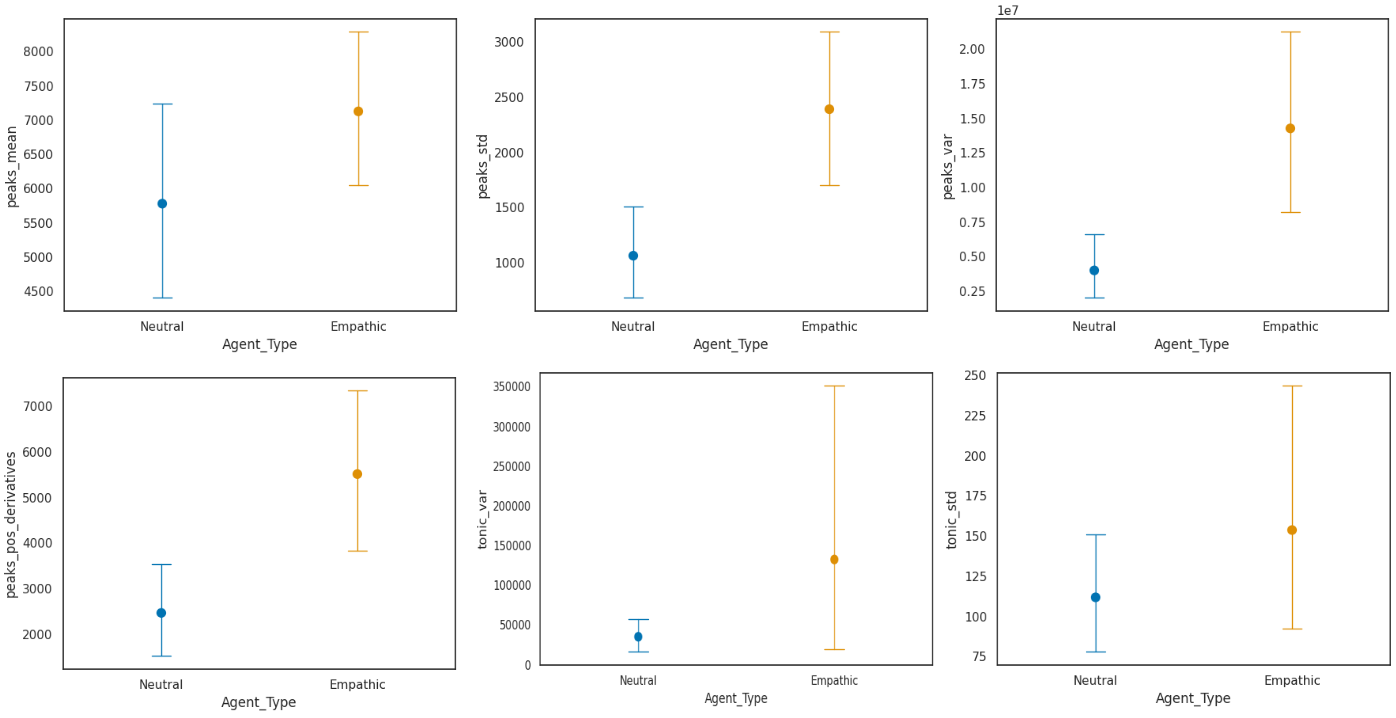}
  \caption{Point-plots of EDA features.}
  \Description{includes point plots of 6 features of EDA which is higher in interaction with empathetic agent}
  \label{fig:EDA}
\end{figure*}

The EDA analysis results revealed that the type of agent had a significant impact on the EDA signals, leading to differences in the statistical characteristics of EDA peaks and the tonic component (Table \ref{tab:EDAPPG}). As depicted in Figure \ref{fig:EDA}, the average amplitude of EDA peaks was higher in interactions with the empathetic agent. A higher EDA peak amplitude indicates a greater increase in skin conductance, which is associated with higher emotional arousal \cite{benedek2010continuous}. Additionally, the higher positive derivatives of EDA peaks observed in interactions with the empathetic agent suggest that the skin conductance is increasing at a faster rate during these interactions, further indicating higher emotional arousal.

\begin{table}[]
\small

 \caption[Statistical result of EDA and PPG features]{The result of statistical analysis for significantly different features for EDA and PPG signal in interaction with different agents}
  \label{tab:EDAPPG}
\centering

\begin{tabular}{|l|l|c|c|c|c|}
\hline
\textbf{Modality}              & \textbf{Variable}       & \textbf{F} & \multicolumn{1}{l|}{\textbf{Dof}} & \multicolumn{1}{l|}{\textbf{Dof.res}} & \textbf{p}      \\ \hline
\multirow{6}{*}{\textbf{EDA}}  & peaks\_var              & 9.12       & 1                                 & 111                                   & \textless 0.004 \\ \cline{2-6} 
                               & peaks\_std              & 9.12       & 1                                 & 111                                   & \textless 0.004 \\ \cline{2-6} 
                               & peaks\_mean             & 5.71       & 1                                 & 111                                   & \textless 0.020 \\ \cline{2-6} 
                               & peaks\_pos\_derivatives & 8.23       & 1                                 & 111                                   & \textless 0.005 \\ \cline{2-6} 
                               & tonic\_var              & 6.78       & 1                                 & 111                                   & \textless 0.020 \\ \cline{2-6} 
                               & tonic\_std              & 6.78       & 1                                 & 111                                   & \textless 0.020 \\ \hline
\multirow{13}{*}{\textbf{PPG}} & HRV\_CVNN               & 11.14      & 1                                 & 103                                   & \textless 0.002 \\ \cline{2-6} 
                               & HRV\_SDNN               & 10.82      & 1                                 & 103                                   & \textless 0.002 \\ \cline{2-6} 
                               & HRV\_RMSSD              & 9.98       & 1                                 & 103                                   & \textless 0.003 \\ \cline{2-6} 
                               & HRV\_SDSD               & 9.92       & 1                                 & 103                                   & \textless 0.003 \\ \cline{2-6} 
                               & HRV\_CVSD               & 9.29       & 1                                 & 103                                   & \textless 0.003 \\ \cline{2-6} 
                               & HRV\_HTI                & 8.89       & 1                                 & 103                                   & \textless 0.004 \\ \cline{2-6} 
                               & HRV\_TINN               & 8.44       & 1                                 & 103                                   & \textless 0.005 \\ \cline{2-6} 
                               & HRV\_MadNN              & 6.25       & 1                                 & 103                                   & \textless 0.020 \\ \cline{2-6} 
                               & HRV\_MedianNN           & 5.48       & 1                                 & 103                                   & \textless 0.030 \\ \cline{2-6} 
                               & HRV\_MCVNN              & 5.16       & 1                                 & 103                                   & \textless 0.030 \\ \cline{2-6} 
                               & HRV\_IQRNN              & 4.76       & 1                                 & 103                                   & \textless 0.040 \\ \cline{2-6} 
                               & HRV\_MeanNN             & 4.13       & 1                                 & 103                                   & \textless 0.050 \\ \cline{2-6} 
                               & HRV\_pNN50              & 4.06       & 1                                 & 103                                   & \textless 0.050 \\ \hline
\end{tabular}

\end{table}

While EDA signals may not directly reflect the valence state \cite{melander2018measuring}, they can serve as an indicator of emotional arousal, excitement, and engagement \cite{kreibig2010autonomic}. In the current study, interactions with the empathetic agent resulted in an increase in skin conductance responses. This could be attributed to a more joyful and engaging experience during interactions with the empathetic agent, leading to heightened emotional arousal.

However, it is important to note that some participants provided post-study verbal feedback indicating feelings of nervousness while interacting with the empathetic agent. They found the concept of empathy emanating from a machine to be peculiar and unappealing. for example, some of them mentioned, "The empathic condition was weird and I could not be able to connect with it. I usually expect a neutral reaction from a stranger or a machine.", or "Empathetic Condition was realistic but weird.". These feelings might have contributed to the observed increase in skin conductance responses during these interactions. 

\subsection{Real-time Expressions}
Throughout the interactions with the DH, three factors were collected to evaluate the emotions expressed by the DH in response to the human participants. These factors focused on assessing the appropriateness of reflected emotions, the expression of emotions, and the timing of these expressions. To understand the impact of the agent type on these factors, a set of one-way repeated measure ANOVA with ART was performed. The result of statistical analysis demonstrated that the agent type significantly affected the appropriateness (F(1, 166) = 10, p \textless 0.002) and timing (F(1, 166) = 6, p \textless 0.01) of the reflected expressions.

\begin{figure*}[h!]
  \centering
  \includegraphics[scale=0.25]{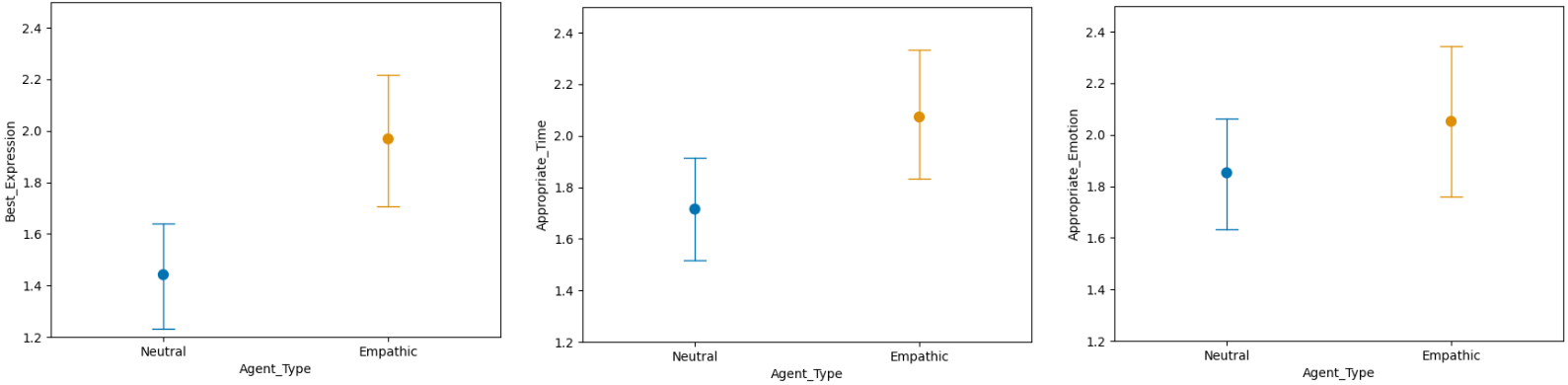}
  \caption{Effect of agent type on the appropriateness of emotion.}
  \Description{3 point plots of the effect of agent type on the appropriateness of emotion. All is higher in interaction with empathetic agent}
  \label{fig:agent_appropriate}
\end{figure*}

In Figure \ref{fig:agent_appropriate}, the point plot illustrates the effect of agent type on the appropriateness of real-time emotions, expressions, and their timing. It is evident from the figure that the empathetic agent tended to use reflected expressions that were more appropriate and at the correct timing. Moreover, the reflected emotion quality was better in the empathetic DH, although this improvement did not reach statistical significance. 

To explore the appropriateness of empathetic DH expressions in different emotional states, a series of two-way repeated measure ANOVA with ART was conducted on participants' ratings in the empathetic condition, with arousal and valence as the conditions. To consider the individual's actual emotional state during the conversation, we mapped the reported arousal and valence using Equations \ref{eq:binary_arousal} and \ref{eq:binary_valence}, respectively. These values served as the emotional state conditions in the statistical analysis.

The results revealed significant differences in the empathetic DH's effectiveness in displaying appropriate emotions, expressions, and timing across different valence levels. The empathetic DH's effectiveness in displaying appropriate emotions and expressions across arousal levels was also significant. However, there were no significant interactions between arousal and valence levels concerning emotion, expression, and timing appropriateness (Table \ref{tab:emotion_appropriate}).

\begin{table}[]
\centering
\caption[Effect of emotional state on the appropriateness of emotion]{Effect of emotional state on the appropriateness of emotion.}
  \label{tab:emotion_appropriate}
\begin{tabular}{|l|l|c|c|c|c|}
\hline
\multicolumn{1}{|c|}{\textbf{Variable}} & \multicolumn{1}{c|}{\textbf{Condition}} & \textbf{F} & \textbf{Df} & \textbf{Df.res} & \textbf{p}      \\ \hline
Appropriate Emotion                     & Arousal                                 & 13         & 1           & 84              & \textless 0.001 \\ \hline
Best Expression                        & Arousal                                 & 6          & 1           & 83              & \textless 0.01  \\ \hline
Best Expression                         & Valence                                 & 6          & 1           & 86              & \textless 0.01  \\ \hline
Appropriate Time                        & Valence                                 & 9          & 1           & 85              & \textless 0.003 \\ \hline
Appropriate Emotion                     & Valence                                 & 12         & 1           & 86              & \textless 0.001 \\ \hline
\end{tabular}

\
\end{table}

\begin{figure}[h!]
  \centering
  \includegraphics[scale=0.19]{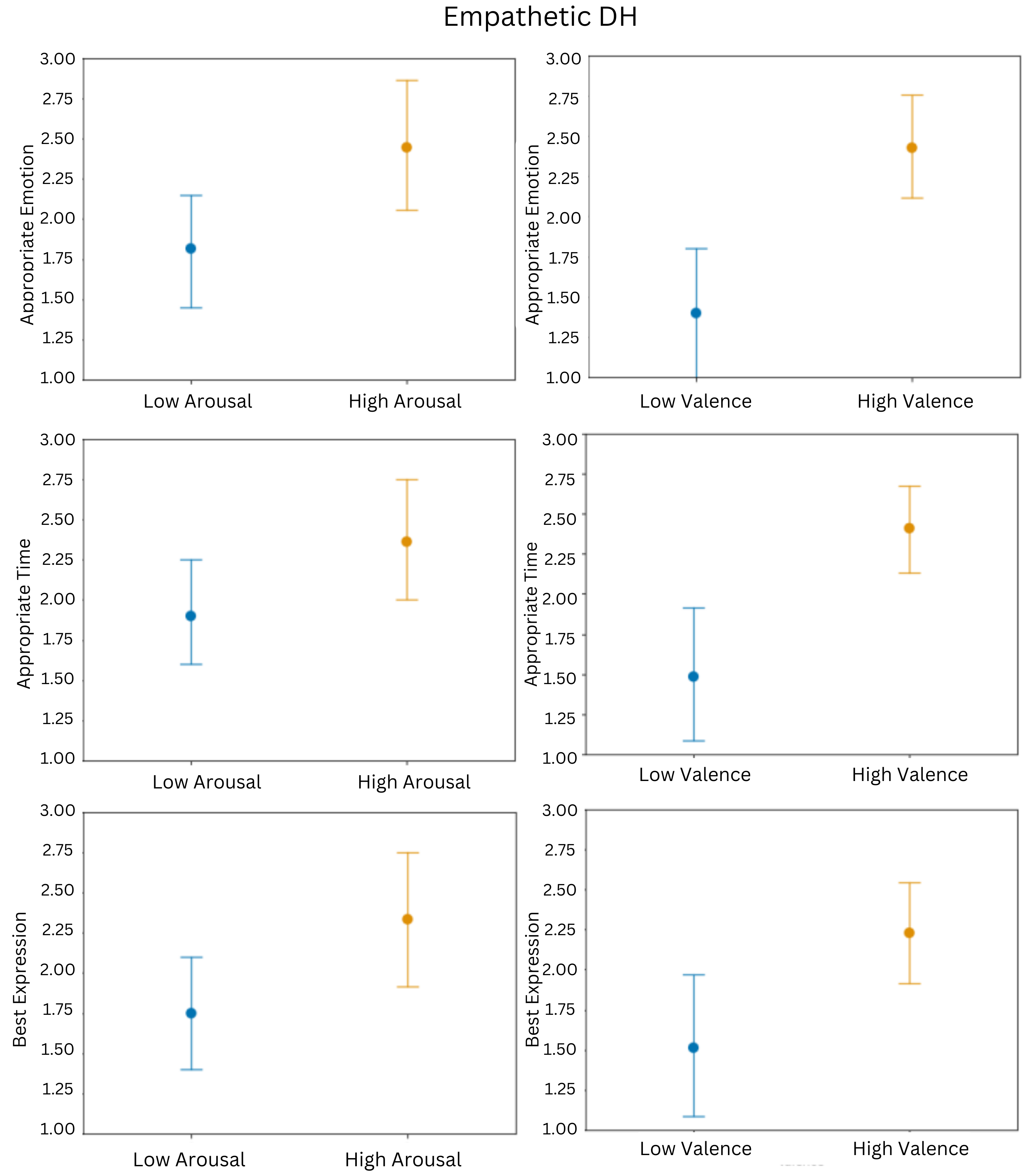}
  \caption[Effect of the emotional state of conversation topic on the appropriateness of emotion]{Point-plots of the effect of emotional state on the appropriateness of emotion, expression, and timing in the empathetic DH.}
  \Description{6 point plots of the emotional state of conversation topic on the appropriateness of emotion. All are higher in interaction with empathetic agent}
  \label{fig:emotion_appropriate}
\end{figure}

The point plots presented in Figure \ref{fig:emotion_appropriate} show that the DH was more successful in expressing positive and high-arousal emotions. One possible explanation could be the model's proficiency in detecting high arousal and valence emotions. Another potential factor might be that the agent struggled to effectively express negative emotions effectively. 

\subsection{Emotion Recognition}\label{subsection:emotion_recognition}
We recorded the detected arousal and valence states using the EEG, PPG, and EDA signals and their fusion in real-time. Then, we compared them with the reported arousal and valence levels. Since the reported labels were based on a 5-point Likert scale, we converted them to low(-1) and high(1) according to Equations \ref{eq:binary_arousal} and \ref{eq:binary_valence}. 

Table \ref{tab:real-time_accuracy} shows the percentage of how well the recognized arousal and valence levels using various modalities matches the reported binary arousal and valence levels. It shows the percentage of correct recognition in real-time using various modalities compared to the reported binary arousal and valence levels. It is evident that the percentage of similarity between the reported emotion and the emotion used for generating real-time expressions and selecting appropriate responses using the fusion method is higher compared to other modalities for both arousal and valence levels. Despite these significant percentages, future improvements in recognition results can be achieved by employing more robust models.
 
\begin{table}[]
\centering
\caption{The percentage of alignment between the recognition of various modalities and the reported arousal and valence levels}
\label{tab:real-time_accuracy}
\begin{tabular}{|l|c|c|}
\hline
                & \multicolumn{1}{l|}{\textbf{Arousal}} & \multicolumn{1}{l|}{\textbf{Valence}} \\ \hline
\textbf{EEG}    & 64.8                                  & 57.1                                  \\ \hline
\textbf{PPG}    & 65.4                                  & 58.1                                  \\ \hline
\textbf{EDA}    & 58.2                                  & 52.7                                  \\ \hline
\textbf{Fusion} & \textbf{69.1}                         & \textbf{57.3}                         \\ \hline
\end{tabular}
\end{table}
\raggedbottom

\subsection{Emotional state}
\begin{figure}[h!]
  \centering
  \includegraphics[scale=0.25]{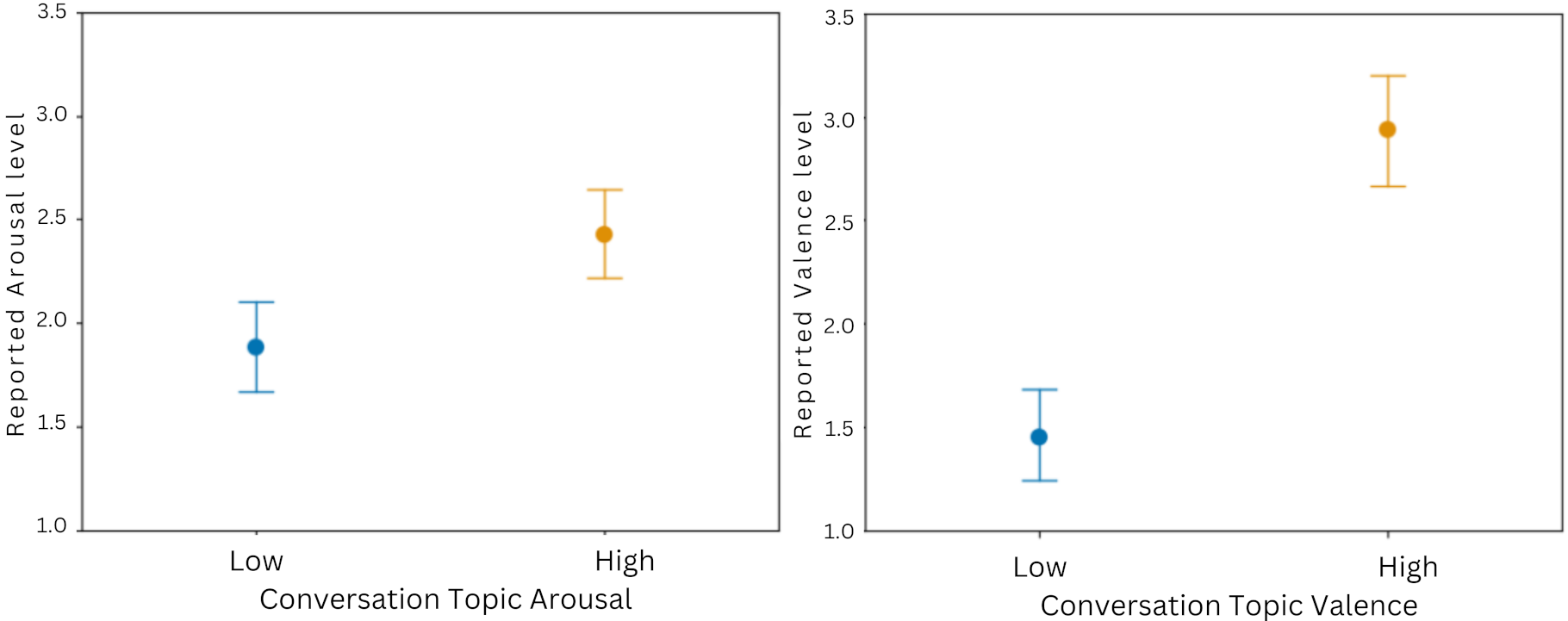}
  \caption{Comparing the reported arousal and valence levels with the emotional state of the conversation topics in order to estimate the effectiveness of inducing emotion. In these two point-plots, the x-axis shows the emotional state (arousal or valence level) of the conversation topic, and the y-axis shows the average of the reported emotional state based on a 5-point Likert scale. It shows a good level of alignment between the conversation topic and emotional state}
  \label{fig:emotion_emotion}
\end{figure}

We conducted an assessment by comparing the reported arousal and valence levels with the emotional state of the conversation topics in order to estimate the effectiveness of conversations in inducing emotions. To achieve this, we utilized a one-way repeated measure ANOVA with ART, which allowed us to examine the influence of the emotional state of the conversation on individuals' emotions. The results indicated that the emotional state of the conversation was significantly affected by the target emotion of the conversation (p-value \textless 0.05). As shown in Figure \ref{fig:emotion_emotion}, the depicted data reveals that the reported arousal and valence ratings correspond closely with the emotional state of the conversation topics. This indicates that the conversations were effective in eliciting emotions.

\raggedbottom

\section{Discussion}
In this paper, we created a real-time emotion recognition application that recognizes emotions using multimodal data. We combined neural and physiological signals to recognize emotion every 5 seconds in real-time and integrated it with a DH from Soul Machines company. The real-time emotions fed into the DH, and DH created its expressions and responses according to the detected emotions empathetically by mirroring the participant's emotional state according to table \ref{tab:reflected_expressions}. Using the multimodal real-time emotion recognition system and the integrated DH, we designed a user study and compared human physiological responses and self-reports in an empathetic and neutral mode.

The results show that people had emotional conversations with neutral and empathetic agents. They felt stronger emotions during their interactions with the empathetic agents. However, based on self-report data, in negative conversations, the empathetic agent was not as successful in understanding or sharing the emotions. This is likely because the agent could not effectively identify and reflect negative emotions. 

The empathetic agent's expressions and responses were more appropriate in positive conversations and at the appropriate time. However, it could not show appropriate expressions at the right time in the negative conversations. The inability to show appropriate expressions at the right time in negative conversations made it difficult for the agent to build rapport and trust with participants. If an agent is not able to effectively respond to negative emotions, it may come across as insensitive or uncaring. Or maybe it is not appropriate to mirror negative emotions like sadness, fear, and anger. It might be more appropriate to have neutral or concerned expressions in these circumstances. This could be explored in future research.

Our analysis of physiological signals also revealed a significantly higher level of HRV and skin conductance responses in interaction with the empathetic agent. These findings show that people experienced more emotional arousal and engagement in interaction with the empathetic agent. They also had better emotion regulation in interaction with the empathetic agent than the neutral agent.

Overall, people enjoyed and engaged more and felt more empathy in interaction with the empathetic DH. This shows that responding based on neural and physiological signals effectively creates empathetic agents and increases the empathy between humans and agents. However, the findings of this study suggest that there is still room for improvement in the development of empathetic agents by improving emotion recognition accuracy.

In this study, we faced some challenges and limitations, as discussed below. We could create more empathetic agents by addressing these shortcomings in the future. 

\begin{itemize}
    \item \textbf{Recognition Accuracy:}
    The level of empathy between humans and the digital human agent was directly influenced by the accuracy of the emotion recognition method. However, achieving a robust model capable of accurately and swiftly detecting emotions using EEG and physiological signals requires further exploration and research. In the future, we can enhance recognition accuracy by using deep learning methods and adopting a more robust fusion strategy. It is essential to also consider the recognition speed while seeking to improve accuracy.

    Incorporating behavioral modalities such as facial expressions and speech recognition could also offer the benefits of both behavioral and physiological modalities. By integrating these modalities, we can potentially enhance the overall performance and effectiveness of the empathetic conversational agent.

\item \textbf{Avatar gender:}
Due to the constraints in participant recruitment, such as the difficulty of the EEG setup, we considered only a female avatar to interact with participants. However, some individuals find it easier to interact with female interviewers, while others are more at ease with male interviewers \cite{brophy1985interactions, tam2020female}. To mitigate the effect of the avatar's gender in future studies, increasing the sample size and considering a male avatar would be beneficial. 

    \item \textbf{Emotional Transition:}
    Different modalities reflect the emotional transition in different time scales. For instance, the EEG signal can swiftly predict emotional transitions, while the PPG signal takes some time to accurately reflect the real emotion, resulting in differing recognition outcomes across modalities. Although this study did not extensively explore emotional transitions, such transitions commonly occur in everyday conversations. Further research is required to investigate the impact of emotion transitions on different modalities and their recognition accuracy to better understand and leverage their potential in empathetic conversational agents.

    \item \textbf{Personality Effect:}
    Individuals with varying personalities may experience differing levels of empathy with the same agent in similar situations. Some individuals might feel uneasy or awkward interacting with a machine and expressing themselves, while others may feel more at ease and relaxed interacting with the agent compared to a human. To further enhance the reliability of the findings, future studies could recruit a larger and more diverse participant pool, thereby reducing the potential influence of personality variations on the results. 

    \item \textbf{Voice Tone}
    The version of the Soul Machines' DH used in this study lacked the capability to control the voice tone to reflect its emotional state. Incorporating an agent that can modulate its voice tone based on the user's emotional state would be beneficial to create a more empathetic interaction in the future.

    \item \textbf{Language model}
    The version of Soul Machines' Digital Human (DH) utilized in this study did not support the use of Large Language Models (LLMs). As a result, the agent's responses were drawn from a predefined database tailored to the detected emotional conditions. In the future, integrating LLMs could significantly enhance the quality of interactions by enabling more dynamic, context-aware, and engaging conversations, thereby improving overall user experience and engagement.

\end{itemize}
\section{Conclusions}
This research aimed to develop and implement a real-time emotion recognition application integrated with a digital human (DH) agent to facilitate empathetic interactions between humans and agents. The novelty lies in the integration of neural and physiological signals into the perception module of conversational agents, enhancing empathetic interactions through real-time physiological and neural emotion recognition. This is one of the first examples of using physiological cues to enable a DH to identify and respond to a user's emotional state empathetically by just relying on neural and physiological cues.

The findings indicate that participants experienced stronger emotions and engagement during interactions with empathetic agents, as evidenced by physiological responses and self-reports. However, challenges and limitations were also encountered, highlighting areas for improvement in future research.

Key areas for enhancement include using large language models (LLMs) for making the conversations and improving the accuracy of emotion recognition methods. LLMs, such as GPT-based models, can generate more natural, context-aware, and engaging dialogue. By incorporating these models, we can simulate human-like conversations that adapt dynamically to user input, making interactions feel more authentic and engaging. 

Advanced deep learning techniques can be used to improve the accuracy of emotion detection. In addition, integrating behavioral modalities such as facial expression analysis and speech recognition enriches the understanding of user emotions. For example, combining data from facial expressions, vocal tone, and textual content can provide a more comprehensive emotional profile. This holistic approach ensures the system can detect and respond to emotions even when one modality is less informative or unavailable. However, we should consider the limitations of each modality in all situations.

Incorporating voice tone modulation capabilities into digital human agents could considerably enhance the empathetic interaction experience, addressing a current limitation observed in the study.

Overall, while the findings demonstrate the potential of real-time emotion recognition and empathetic digital agents, ongoing research is essential to address the identified challenges and further refine the development of empathetic conversational agents. By addressing these limitations, future studies can contribute to the advancement of human-agent interaction and foster more empathetic and meaningful connections between humans and artificial entities.

\printbibliography

\end{document}